\documentclass[useAMS]{mn2e}
\usepackage{epsfig}
\usepackage{amsmath}
\usepackage{pifont}

\title[Well--correlated diffuse bands: 6196 and 6614 \AA]
{Observational analysis of the well--correlated diffuse bands: 6196 and 6614 \AA\
\thanks{This paper includes data gathered with FEROS, HARPS, UVES spectrographs under programs 073.C-0337(A), 079.C-0597(A), 082.C-0566(A).
 }}

\author[Kre{\l}owski et al.]
{J. Kre{\l}owski$^{1}$,  G.A. Galazutdinov$^{2,3,4}\thanks{runizag@gmail.com}$, A. Bondar$^5$ \& Y. Beletsky$^6$\\
$^{1}$Center for Astronomy, Nicholas Copernicus University,
Gagarina 11, Pl-87-100 Toru{\'n}, Poland \\
$^{2}$Instituto de Astronomia, Universidad Catolica del Norte, Av. Angamos 0610, Antofagasta, Chile\\
$^{3}$Pulkovo Observatory, Pulkovskoe Shosse 65, Saint-Petersburg, 196140, Russia\\
$^{4}$Special Astrophysical Observatory, Nizhnij Arkhyz, 369167, Russia \\
$^{5}$International Center for Astronomical and Medico-Ecological Research, Zabolotnoho Str. 27, Kiev, 03187, Ukraine \\
$^{6}$Las Campanas Observatory, Carnegie Observatories, Casilla 601, La Serena, Chile.}

\begin{document}

\date{Accepted . Received ; in original form }

\pagerange{\pageref{firstpage}--\pageref{lastpage}} \pubyear{2002}

\maketitle

\label{firstpage}

\begin{abstract}
We confirm, using spectra from seven observatories, that the diffuse
bands 6196 and 6614 are very tightly correlated. However, their
strength ratio is not constant as well as profile shapes.
Apparently the two interstellar features do not react in unison to
the varying physical conditions of different interstellar clouds.
\end{abstract}


\section{Introduction}

A very recent paper by Oka et al. (2013) demonstrates that
profiles of the still unidentified diffuse interstellar bands
(DIBs), in particular the very well--correlated ones near 6196 and
6614~\AA, may not change in unison (see their Figs. 1 and 2).
The rotational temperatures of the polar molecules CH and CH$^+$,
found in Herschel 36, are as high as about 15K. In this special
object the 6614 profile shows a very pronounced extended red wing
while the latter is barely seen in 6196.
 There are several possible explanations for
the origin of red wing seen in DIB 6614: Bernstein et al. (2015)
offer the hypothesis of accidentally overlapping diffuse bands,
Oka et al. (2013) consider a single rotational contour where
high rotational transitions being pumped by the nearby irradiation from Her 36.
Lastly, Marshall et al (2015) suggest that "...the extended
red tails would arise principally from vibrational hot bands that
are red-shifted with respect to the origin band..."

The possible influence of
the rotational temperatures of centrosymmetric species (C$_2$) on
some DIB profiles was already suggested by Ka{\'z}mierczak et al.
(2010). This result received a support very recently
(Gnaci{\'n}ski \& Kre{\l}owski (2014); some DIB profile widths
have been shown as related to the rotational temperature of H$_2$
-- the homonuclear molecule.

\begin{table*}
\caption{Equivalent widths of the considered DIBs in spectra of
our targets. Origin of spectra is marked as follows: b - BOES, f - FEROS,  h - HARPS, M - McDonald, t - Terskol, u - UVES.
Weighted averages of the same target data are boldfaced.
}
\label{ew}
\begin{tabular}{lrr lrr lrr}
\hline
Star         &  EW(6196)     &    EW(6614)  &  Star    &  EW(6196)    &    EW(6614)  &  Star     &  EW(6196)      &    EW(6614)  \\
\hline
bd59456b     & 52.8$\pm$2.3 & 224.2$\pm$6.0 & 144470f      & 15.8$\pm$1.3 &  60.4$\pm$2.7 &      154445u  & 24.0$\pm$0.7 & 103.9$\pm$1.3 \\
bd60594b     & 40.2$\pm$1.7 & 171.1$\pm$5.1 & 144470h      & 18.6$\pm$3.1 &  57.8$\pm$5.2 &      157038u  & 46.8$\pm$0.6 & 179.0$\pm$2.7 \\
bd404220b    & 89.5$\pm$2.5 & 300.7$\pm$5.2 &       &{\bf 16.2$\pm$1.0}&{\bf 59.8$\pm$1.1}&     161056u  & 38.0$\pm$1.5 & 161.0$\pm$3.6 \\
bd404227t    & 84.6$\pm$4.6 & 327.4$\pm$9.9 & 145037u      & 72.6$\pm$2.1 & 244.0$\pm$6.0 &      163800h  & 27.8$\pm$2.5 & 118.2$\pm$2.5 \\
CD-324348u   & 82.4$\pm$0.8 & 344.6$\pm$2.5 & 145502f      & 15.0$\pm$0.6 &  58.0$\pm$1.9 &      166734u  & 97.4$\pm$1.1 & 417.1$\pm$3.7 \\
CygOB27b     & 90.0$\pm$4.1 & 344.8$\pm$9.9 & 147165h      & 18.0$\pm$0.7 &  60.1$\pm$1.8 &      167971u  & 62.5$\pm$1.0 & 250.9$\pm$3.5 \\
CygOB28ab    & 94.9$\pm$4.0 & 348.7$\pm$7.0 & 147165u      & 17.6$\pm$0.6 &  58.0$\pm$1.5 &      168607u  & 97.3$\pm$1.2 & 387.4$\pm$3.8 \\
CygOB211b    &112.0$\pm$5.7 & 380.0$\pm$7.8 &       &{\bf 17.8$\pm$0.2}&{\bf 58.9$\pm$1.0}&      168625u  &100.5$\pm$1.0 & 434.2$\pm$2.2 \\
CygOB212b    &116.8$\pm$9.8 & 386.4$\pm$9.9 & 147889h      & 46.2$\pm$2.0 & 179.5$\pm$3.4 &      169454u  & 59.9$\pm$2.8 & 196.6$\pm$3.3 \\
15785b       & 46.7$\pm$1.2 & 221.3$\pm$4.8 & 147889u      & 45.4$\pm$0.9 & 181.6$\pm$1.5 &      169454M  & 59.3$\pm$1.2 & 194.8$\pm$2.0 \\
23180t       & 15.3$\pm$1.6 &  61.8$\pm$5.1 &      &{\bf 45.5$\pm$0.3}&{\bf 181.3$\pm$0.8}&          &{\bf 59.4$\pm$0.2}&{\bf 195.3$\pm$0.8}\\
24398b       & 13.9$\pm$1.0 &  58.1$\pm$4.6 & 147933h      & 15.9$\pm$1.0 &  62.7$\pm$2.7 &      170740u  & 28.8$\pm$1.1 & 116.9$\pm$1.7 \\
24912t       & 22.7$\pm1.3$ &  95.3$\pm$3.2 & 148184h      & 12.5$\pm$0.7 &  42.1$\pm$2.0 &      179406h  & 20.0$\pm$0.9 &  94.2$\pm$2.2 \\
27778M       & 10.6$\pm$0.8 &  45.5$\pm$1.9 & 148184u      & 13.2$\pm$0.6 &  40.5$\pm$1.7 &      183143m  & 96.0$\pm$1.3 & 358.5$\pm$2.8 \\
34078M       & 24.9$\pm$1.2 &  66.0$\pm$3.1 &       &{\bf 12.9$\pm$0.3}&{\bf 41.2$\pm$0.8}&      184915m  & 20.4$\pm$1.2 &  77.3$\pm$2.1 \\
34078t       & 22.0$\pm$1.2 &  54.0$\pm$3.1 & 148379u    & 42.8$\pm$0.8 & 159.0$\pm$2.5   &        184915t  & 17.8$\pm$2.7 &  73.2$\pm$4.2 \\
34078u       & 25.4$\pm$2.1 &  65.9$\pm$4.0 & 149408u    & 45.9$\pm$1.4 & 177.0$\pm$2.6   &           &{\bf 20.0$\pm$1.0}&{\bf 76.5$\pm$1.6}\\
      &{\bf 23.7$\pm$1.1}&{\bf 61.4$\pm$4.1}& 149757h    & 10.5$\pm$0.7 &  41.8$\pm$1.9   &        194839t  & 66.1$\pm$3.9 & 197.5$\pm$9.0 \\
63804u       & 87.6$\pm$1.7 & 332.0$\pm$3.5 & 149757u    & 11.2$\pm$0.8 &  40.7$\pm$1.5   &        203532M  & 14.0$\pm$0.8 &  58.4$\pm$1.1 \\
73882u       & 17.1$\pm$0.9 &  48.1$\pm$1.8 &       &{\bf 10.8$\pm$0.3}&{\bf 41.1$\pm$0.5}&     204827b  & 41.2$\pm$2.6 & 173.1$\pm$4.0 \\
78344u       & 76.6$\pm$2.1 & 281.4$\pm$4.6 & 151932u    & 31.6$\pm$0.9 & 121.8$\pm$2.1   &        207198t  & 33.0$\pm$2.0 & 125.2$\pm$3.7 \\
80077u       & 80.2$\pm$1.2 & 291.3$\pm$1.9 & 152003u    & 32.9$\pm$2.0 & 105.8$\pm$2.9   &        210839t  & 32.9$\pm$0.8 & 145.5$\pm$3.2 \\
110432u      & 17.0$\pm$0.3 &  74.4$\pm$1.6 & 152233h    & 22.6$\pm$1.2 &  78.3$\pm$2.4   &        226868b  & 49.0$\pm$3.8 & 180.2$\pm$4.9 \\
114213u      & 38.9$\pm$0.9 & 145.2$\pm$3.7 & 152233M    & 22.7$\pm$1.1 &  79.3$\pm$2.1   &        228712b  & 57.4$\pm$2.7 & 181.4$\pm$3.6 \\
115842u      & 36.8$\pm$1.1 & 151.0$\pm$2.0 &       &{\bf 22.7$\pm$0.1}&{\bf 78.9$\pm$0.5}&     228779b  & 71.8$\pm$2.1 & 285.9$\pm$6.0 \\
125241u      & 57.1$\pm$0.8 & 224.3$\pm$3.4 & 152235h    & 36.3$\pm$1.1 & 132.5$\pm$2.6   &        229059t  & 63.8$\pm$4.8 & 231.4$\pm$7.5 \\
142468u      & 56.4$\pm$0.8 & 220.2$\pm$3.4 & 152236u    & 39.3$\pm$1.2 & 129.0$\pm$2.1   &        254577b  & 51.7$\pm$2.2 & 206.0$\pm$4.5 \\
144217h      & 13.2$\pm$0.8 &  40.9$\pm$2.0 & 152249h    & 20.0$\pm$1.0 &  86.4$\pm$2.7   &        278942b  & 32.0$\pm$2.1 & 166.0$\pm$4.6 \\
144217u      & 11.1$\pm$0.3 &  45.6$\pm$1.5 & 152249u    & 19.2$\pm$0.6 &  81.5$\pm$1.9   &        Hersch36 & 40.6$\pm$4.1 & 151.3$\pm$9.9 \\
      &{\bf 11.4$\pm$0.7}&{\bf 43.9$\pm$2.3}&       &{\bf 19.4$\pm$0.4}&{\bf 83.1$\pm$2.3}&                 &              &               \\
144218h      & 14.1$\pm$3.1 &  57.4$\pm$8.6 &   154368u  & 33.8$\pm$1.9 & 150.0$\pm$3.3   &                 &              &               \\
\hline
\end{tabular}
\end{table*}

The pair of the diffuse bands near 6196 and 6614~\AA\ was found as
very well correlated by Moutou et al. (1999). The intensities of
the two DIBs were found as correlated with the Pearson's
coefficient as high as 0.97. The first guess could thus be that
6196 and 6614 may form a ``family'' sharing a common carrier.
However, the very high resolution analysis of the profiles of the
two DIBs created some doubts (Galazutdinov et al., 2002). Their
Fig. 4 demonstrated that while EW's of 6196 and 6614 do correlate
quite tightly, their FWHM's do not. Apparently the profiles of
both DIBs do not react in the same fashion to the varying physical
conditions inside interstellar clouds. This fact is also confirmed
by differing shapes of both DIB profiles (Galazutdinov, Lo Curto
\& Kre{\l}owski 2008).

 The nearly perfect correlation of the two bands was confirmed by
McCall et al. (2010) which is better than any other pair of DIBs.
Nevertheless, the authors emphasized the need to explain the very different profiles of
the 6196 and 6614 diffuse bands if they are of the same origin. Indeed, profile of the former one
 is much more symmetric and narrow than that of the latter.

Naturally, it was the very first guess that trying to divide
diffuse bands into sets of features carried by the same molecules
we should put into one ``family'' the features which are really
well correlated (e.g. Kre{\l}owski, Schmidt \& Snow (1997)).
However, the above mentioned papers expressed some doubts on
whether the two nearly perfectly correlated DIBs share the same
carrier. The profiles are of very different widths and these
widths vary but not in the same way. It is thus important to check
whether the observed scatter, seen despite a very high correlation
coefficient, follows just measurement errors or has some physical
background.

It is commonly believed that the DIB carriers are some complex
molecules. The most popularly considered molecular species, being
candidates for DIB carriers, are chain molecules, based on a
carbon skeleton. They have been observed in star forming regions
but the attempts to find them in translucent clouds failed
(Motylewski et al. 2000). Another candidature is that of
polycyclic aromatic hydrocarbons (PAHs). A kind of their mixture
is believed as the carrier of a few emission infrared bands.
However, the attempts to identify any specific PAH -- this is
possible only in the near UV range -- failed as well (Salama et
al. 2011; Gredel et al. 2011). Another possibility -- fullerenes
-- seems to be restricted to a few DIBs only, despite this species
were discovered in star forming regions (Cami et al. 2010).
Recently Campbell et al. (2016) reported
the identification of C$_60^+$ near infrared features in spectra of relatively cool object (B7Ia) HD\,183143.
Earlier attempts to identify C$_60^+$, e.g. by Jenniskens et al. (1997), Galazutdinov et al. (2000) pointed out
that the lack of analysis of spectra of differentiated spectral types makes the identification of C$_60^+$
uncertain because the interstellar features (especially 9633) may be severely contaminated with stellar lines,
being specific in each case. Unfortunately, this critical issue have not been addressed by Campbell et al.

\section{Observations}

The UVES spectra of our targets were collected at the 8-m UT2
telescope (Paranal Observatory, ESO, Chile). We used our own
observations as well as those available from the public ESO archive.
The spectra were carried out in two standard modes DIC1(346+580)
and DIC2(437-860) covering the whole wavelength range $\sim$3050 -
$\sim$10400 \AA\AA\, some of them with a gap between $\sim$5770 -
$\sim$5830 \AA\AA\, due to the inaccurate set-up of the optical
elements of spectrograph.

The applied slit width was 0.4 and 0.3 arcsecond for the blue and
red branches of the spectrograph respectively. These widths
satisfy to the 2 pixel criteria for the slit image projection to
the corresponding CCD cameras, providing the highest possible
resolving power $\sim$80,000 and $\sim$110,000 for the blue and
red spectrograph branches respectively.

Several spectra used in this investigation were acquired using the
HARPS spectrograph fed with the 3.6\textendash m LaSilla
telescope. This instrument offers a very high resolution
(R=115,000); the wavelength range is 3,800 -- 6,900~\AA\ divided
into 72 orders.

Additional spectra  were collected with the aid of the
Magellan/Clay telescope at the Las Campanas Observatory (Chile)
using the MIKE spectrograph (Bernstein et al. 2003) with a
0.35$\times$5 arcs slit. The resulting high resolution spectrum
(R$\sim$65,000) is an average of several individual exposures,
achieves the high S/N ratio ($\sim$ 1000) and covers the broad
wavelength range $\sim$3600 - $\sim$ 9400 \AA\AA) with the lack of
gaps between spectral orders.

A big portion of the data was collected at the ICAMER (North
Caucasus, Russia) observatory, with the aid of the coud\'e echelle
spectrograph MAESTRO (Musaev 1999)
fed by a 2\textendash m telescope. These spectra may be obtained
in two operating modes of the instrument, with resolving powers
respectively of 45,000 and 120,000. In this paper we used only the
high (R=120,000) resolution.

We have measured our selected diffuse bands' equivalent widths
also in the spectra collected at the McDonald Observatory in 1993
by one of us (JK). The Sandiford echelle spectrograph allows the
resolution as high as R=60,000 and the wavelength range 5,650 --
7,000~\AA\ is divided into 27 orders.

Several targets have been observed using the Feros spectrograph,
fed with the 2.2\textendash m telescope. The resolution of this
instrument is R=48,000 and the spectral range covers 3,700 --
9,000~\AA. The latter is divided into 37 orders.

Lastly, our sample includes seven spectra obtained through the
fibre\textendash fed echelle spectrograph BOES (Kim et al., 2007)
installed on the 1.8-m telescope at the Bohyunsan Observatory
(South Korea). In this case the resolution may be either 30,000;
45,000 or 90,000; the wavelength range is always 3,800 --
10,000~\AA\ divided into 76 orders.

All spectra were processed and measured in a standard way using
both {\sevensize IRAF} (Tody 1986) and our own
{\sevensize DECH}\footnote[1]{http://gazinur.com/DECH-software.html} codes.

\section{Results}

\begin{figure}
\includegraphics[angle=270,width=8.5cm]{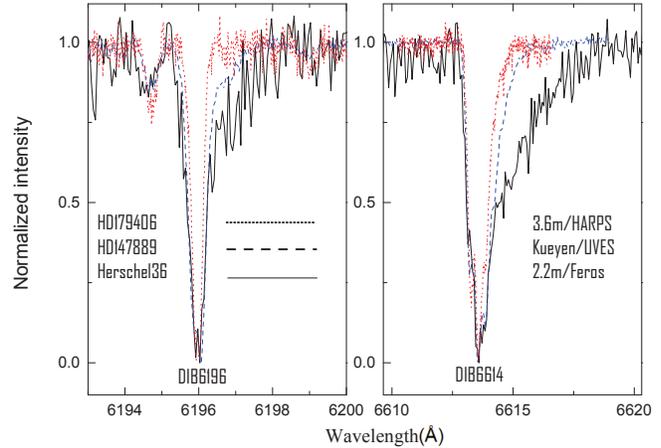}
\caption{The extended red wings of the considered DIBs growing in
unison with T$_{rot}$ of identified species.}
\label{erw03}
\end{figure}

Oka (2013) demonstrated that profiles of both 6196 and 6614
(especially the latter) may change seriously while the rotational
temperatures of CH or CH$^+$ are extraordinarily high, forming the
extended red wings. Our sample of high S/N ratio targets shows
that similar extended red wings may be observed in relatively many
objects though not as evident as in Herschel 36. We have selected
three spectra of HD179406, HD147889 (mentioned as differing
seriously in T$_{rot}$ of C$_2$ by Ka\'zmierczak et al. 2010) and
Herschel 36 to compare the DIB profiles. The result is shown in
Fig. \ref{erw03}. Apparently HD179406 is the case of low
rotational temperatures of molecular species and HD147889 is
somewhere in between of HD179406 and Herschel 36. We would like to
draw attention to a comparison of two spectra of HD147889 and
HD179406: one from HARPS and one from UVES -- both of very high
S/N ratio. The DIB profiles are not identical in both these
sources. Apparently the extended red wings develop with the
growing rotational temperatures of molecular species. In the case
of 6196 the DIB core is broadened when T$_{rot}$ grows but the
extended red wing is relatively shallow while in the case of 6614
it is very pronounced.

Our measurements of the equivalent widths of both DIBs in our
sample of high S/N ratio spectra are given in Table 1.
Equivalent width error estimations are based on use of  equation 7
from Vollmann \& Eversberg (2006). The method neglects possible continuum
placement errors, though hard to expect sufficient incorrectness in case of relatively narrow DIBs like 6196.

\begin{figure}
\includegraphics[angle=270,width=8.5cm]{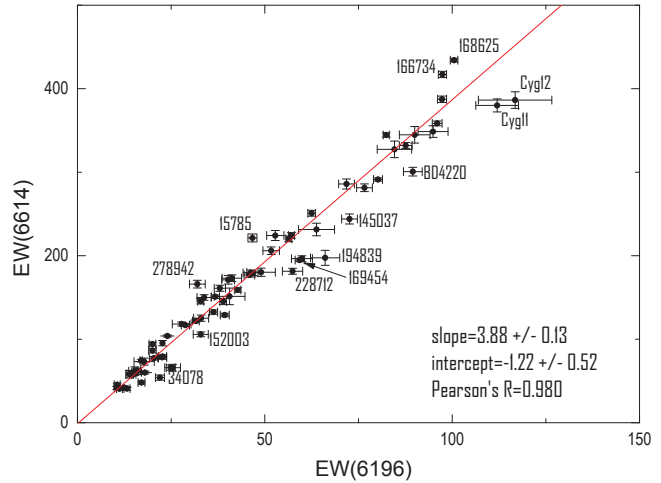}
\caption{The tight relation between 6196 and 6614 diffuse bands. A
few departing points are labeled with the stellar HD or BD
numbers. For HD34078 we plotted measurements from UVES, MIKE and
Maestro. Linear fit done with including errors for both coordinates, see Fasano \& Vio(1988).
} \label{corr}
\end{figure}

It is important to emphasize that the extended red wings are incorporated into the
integrated profiles if present. This set of data is especially
reliable as we have selected from our database only the best
spectra, i.e. with S/N ratio exceeding 250 and where the DIB
profiles are not contaminated with stellar lines. Our sample
contains thus 80 spectra; some of the targets have been observed
using more than one instrument. We consider the fact that
equivalent widths, measured using different instruments, coincide
inside the one sigma errors, as very important. They prove that
all measurements can be repeated with the same result. The
correlation plot is demonstrated in Fig. \ref{corr}.

Fig. \ref{corr} clearly supports the conclusions of the above
mentioned papers -- that the correlation between both DIBs is
nearly perfect. However, it seems interesting to compare the
departing points and so -- to check whether they really do not
follow the main stream because of the measurements' errors or the
strength ratios of 6196 and 6614 in these targets are really
different.

\begin{figure}
\includegraphics[angle=0,width=8.5cm]{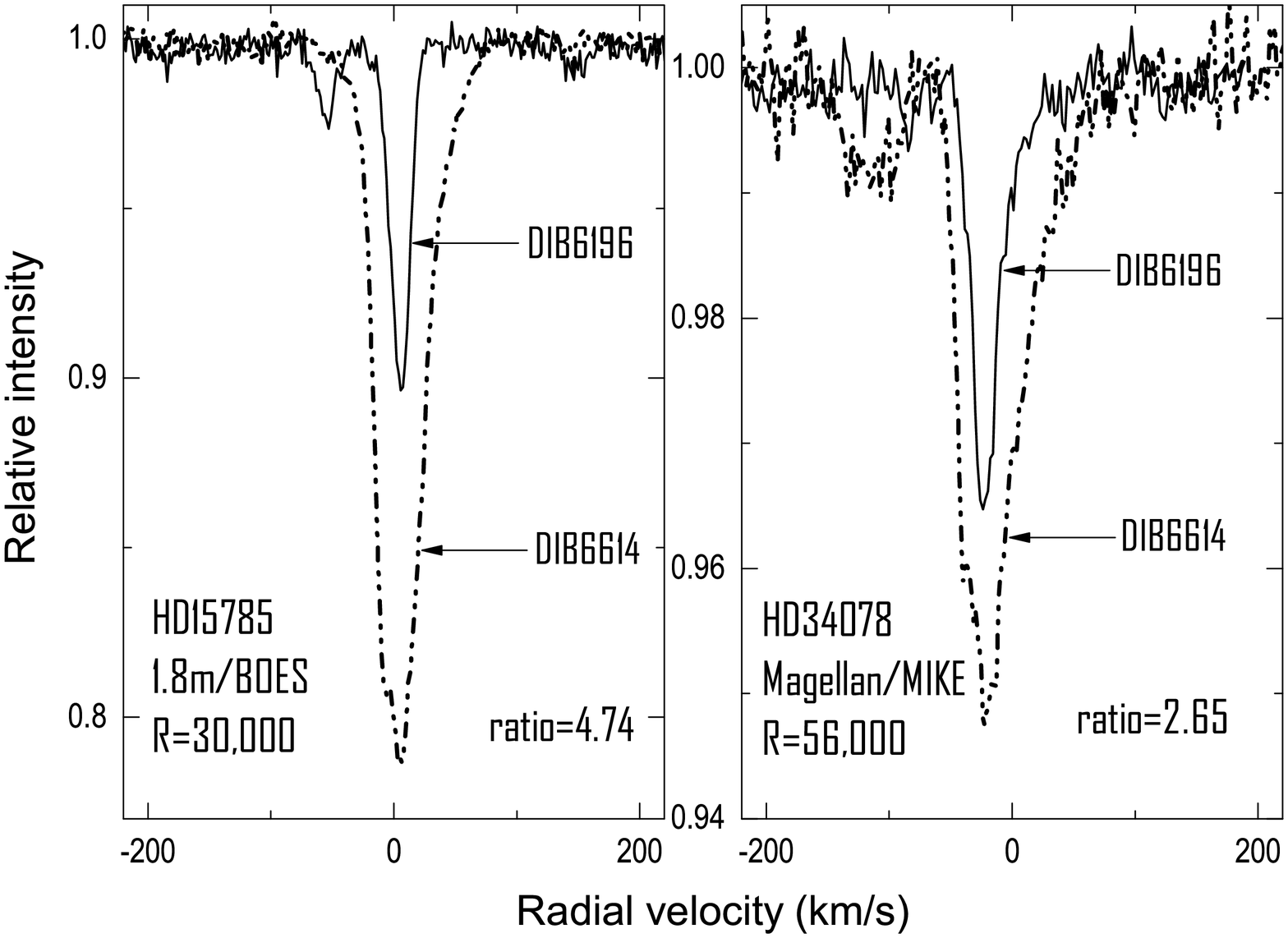}
\caption{The profiles of both DIBs plotted in the radial velocity
scale for the two special targets: HD34078 (AE Aur) and HD15785.
The different strength ratio follows the scatter seen in Fig.
\ref{corr}.}
\label{ratio}
\end{figure}

There are some evidently departing points which represent
HD34078 (weak DIBs but three different spectra) and HD15785 where
both DIBs are strong and the BOES spectrum is of very high
quality. The profiles of both diffuse bands are shown in the
radial velocity scale Fig. \ref{ratio}. Another departing star - HD169454, is
represented by two points: one based on UVES and one -- on MIKE
spectra, both of very high quality.

Fig. \ref{ratio} demonstrates that the strength ratio of the two
DIBs under consideration really varies from target to target.
Apparently there are two different carriers causing these DIBs
though they must be related in a way -- their abundance ratios are
similar but not identical. This seems to support the conclusions
of Galazutdinov et al. (2002) that the FWHM's of the DIBs do not correlate
and the recent result of Oka et al. (2013) which demonstrates
evident extended red wing in 6614 while the latter is barely seen
in 6196. This extended red wing seems to appear in HD34078 and
thus the object may be a transition one between typical objects
with low rotational temperatures of molecules and the extreme
object -- Herschel 36. It is worth to mention that the T$_{rot}$ of the
CN molecule in this object (HD34078) is as high as 4.0K
(Kre{\l}owski et al., 2012). It is interesting that the point, representing Herschel 36 in Fig.
\ref{corr} is situated exactly at the average relation. The
observed 6196/6614 strength ratios seemingly do not depend on
rotational temperatures of identified molecular species.

\begin{figure}
\includegraphics[angle=0,width=8.5cm]{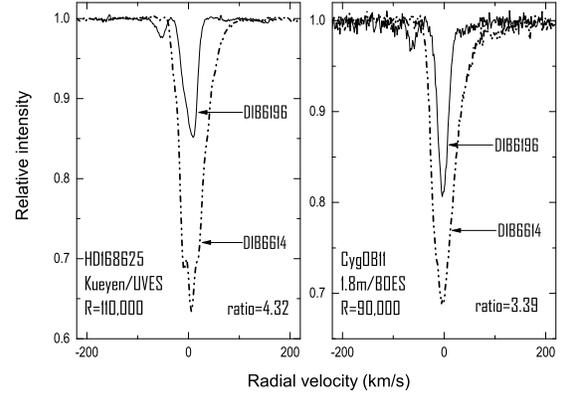}
\caption{The same as in Fig. \ref{ratio} but for Cyg OB2 11 and
HD168625. The different strength ratio in both spectra is shown
beyond a doubt. Apparently the scatter, seen in Fig. \ref{corr}
does not follow measurements' errors -- it reflects different
physical conditions.}
\label{ratio_01}
\end{figure}

\begin{figure}
\includegraphics[angle=0,width=8.5cm]{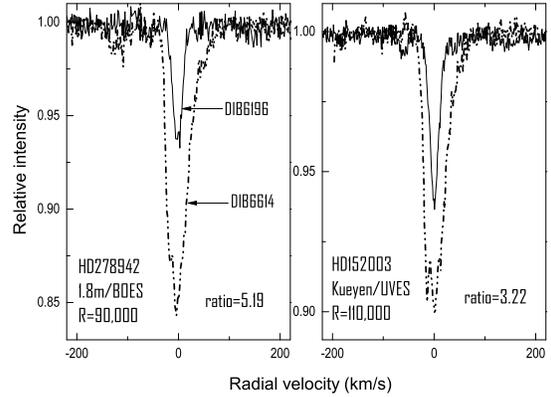}
\caption{The same as in Fig. \ref{ratio} but for HD278942 and
HD152003. The different strength ratio in both spectra is once
again shown beyond a doubt. } \label{ratio_pl}
\end{figure}

Fig. \ref{corr} allows to select other pairs of targets situated
at both sides of the average relation. We also demonstrate the same
as in Fig. \ref{ratio} effect for Cyg OB2 11 and HD168625 and for
HD278942 and HD152003 -- Fig. \ref{ratio_01} and Fig.
\ref{ratio_pl}. The plots clearly support the conclusion that the
two DIBs' ratios really differ from object to object. Apparently
the physical scatter is not related to the changes of DIB
profiles. The profiles apparently depend on rotational
temperatures of the carriers while the strength ratios -- on their
abundances.

\section{Conclusions}

Our analysis of the 6196 and 6614 \AA\ diffuse bands leads to the
following conclusions:
\begin{itemize}
\item
profiles of both diffuse bands change
with the rotational temperature of identified species; the growing
T$_{rot}$ results with the extended red wings
\item
both considered DIBs correlate almost perfectly (with the Pearsons
correlation coefficient 0.98); the average 6614/6196 strength
ratio being 3.88$\pm$0.13
\item
the observed scatter is not a result of measurement's errors,  i.e. the difference is higher than uncertainty of individual measurements; the
6614/6196 strength ratios are proved to be variable
\item
the variable strength ratio is not related to the observed profile
changes; apparently both DIBs are caused by different carriers but
their abundances are pretty closely related.
\end{itemize}

Apparently the very first idea, expressed by Kre{\l}owski \&
Walker (1987) and by Kre{\l}owski \& Westerlund (1988), that all
DIBs do not share the same carriers but can be divided into
several ``families'' cannot be helpful while trying to identify
DIB carriers. An analysis of DIB profiles, pioneered by Westerlund
\& Kre{\l}owski (1988) is necessary as well. As yet it was not
possible to find any pair of DIBs being certainly carried by the
same species. Seemingly all stronger DIBs which can be reliably
measured in statistically meaningful samples of objects are
carried by a different molecule each.

\section*{Acknowledgments}
JK acknowledges the financial support of the Polish National
Center for Science during the period 2016 - 2018 (grant
UMO-2015/17/B/ST9/03397). GAG thanks the Russian Science Foundation (project 14-50-00043, the Exoplanets program) for
support of observational and interpretational parts of this study.

\label{lastpage}
\end{document}